\documentclass[a4paper, 12pt]{scrartcl}

\usepackage[utf8x]{inputenc}
\usepackage[T1]{fontenc}

\usepackage[english]{babel}

\makeatletter
\DeclareOldFontCommand{\rm}{\normalfont\rmfamily}{\mathrm}
\DeclareOldFontCommand{\sf}{\normalfont\sffamily}{\mathsf}
\DeclareOldFontCommand{\tt}{\normalfont\ttfamily}{\mathtt}
\DeclareOldFontCommand{\bf}{\normalfont\bfseries}{\mathbf}
\DeclareOldFontCommand{\it}{\normalfont\itshape}{\mathit}
\DeclareOldFontCommand{\sl}{\normalfont\slshape}{\@nomath\sl}
\DeclareOldFontCommand{\sc}{\normalfont\scshape}{\@nomath\sc}
\makeatother

\usepackage[a4paper,  centering, bindingoffset=0mm, inner=25mm, outer=25mm, top=26mm, bottom=36mm, heightrounded]{geometry}

\usepackage{graphicx}
\usepackage{float}
\usepackage{cite}

\usepackage{amsmath}	
\usepackage{amssymb}
\usepackage{amsfonts}
\usepackage{mathrsfs}
\usepackage{mathtools}
\usepackage{amsthm}

\usepackage{braket}
\usepackage{slashed}

\usepackage{booktabs}
\usepackage{longtable}	
\usepackage{multirow}

\usepackage{epsf}
\usepackage{epsfig}	
\usepackage{epstopdf}
\usepackage[font=small,labelfont=bf, width=.95\textwidth]{caption}

\usepackage{verbatim}
\usepackage{textcomp}
\usepackage{enumitem}

\usepackage[toc]{appendix}

\usepackage{layout}

\usepackage{microtype}
\usepackage{bookmark}

\usepackage{color}
\usepackage[]{xcolor}

\newcommand{\dd}{\mathrm{d}}

\newcommand{\ii}{\mathrm{i}}

\makeatletter
\newcommand{\subalign}[1]{%
  \vcenter{%
    \Let@ \restore@math@cr \default@tag
    \baselineskip\fontdimen10 \scriptfont\tw@
    \advance\baselineskip\fontdimen12 \scriptfont\tw@
    \lineskip\thr@@\fontdimen8 \scriptfont\thr@@
    \lineskiplimit\lineskip
    \ialign{\hfil$\m@th\scriptstyle##$&$\m@th\scriptstyle{}##$\crcr
      #1\crcr
    }%
  }
}
\makeatother

\definecolor{unamblue}{cmyk}{1 0.79 0.12 0.59}

\usepackage[]{hyperref}
\hypersetup{
    colorlinks=true,%
    citecolor=unamblue,%
    filecolor=unamblue,%
    linkcolor=unamblue,%
    urlcolor=unamblue,
    bookmarksnumbered=true,     
    bookmarksopen=true,         
    bookmarksopenlevel=1,       
    pdfstartview=Fit,           
    pdfpagemode=UseOutlines,
    pdfpagelayout=TwoPageRight
}

\usepackage{graphicx}
\usepackage{tikz}
\usepackage{slashed}
\usepackage{epstopdf}
\usepackage{verbatim}	
\usepackage{blkarray}

\title{\Huge A scattering amplitudes approach to hard thermal loops \\[1\baselineskip]}

\author{Leonardo de la Cruz$^{a,b}$\\
\\
\emph{\normalfont\normalsize ${}^a$Dipartamento di Fisica e Astronomia, Universit\'a di Bologna e INFN Sezione di Bologna}\\
\emph{\normalfont\normalsize via Irnerio 46, 	I-40126 Bologna, Italy}	
\\
\emph{\normalfont\normalsize ${}^b$Higgs Centre for Theoretical Physics, School of Physics and Astronomy,}\\
\emph{\normalfont\normalsize The University of Edinburgh,}\\
\emph{\normalfont\normalsize Edinburgh EH9 3FD, Scotland, UK} 
}
\date{%
  $\,$
    \\[2\baselineskip]
    \normalfont\normalsize%
      \parbox{0.8\linewidth}{%
{\bf \sf Abstract}. 
Inspired by recent progress on classical limits of scattering amplitudes, we show that hard thermal loops can be obtained from classical limits of off-shell currents.
The classicality of hard thermal loops is made manifest by associating classical wavenumbers to soft particles. 
We compute the classical limit of these currents in QED, QCD and gravity. Our proposal does not involve the introduction of ghosts. 
}}

\usepackage{tikz}
\usepackage{ytableau}
\usepackage{tikz-feynman}
\tikzfeynmanset{warn luatex=false}

\newcommand{\cc}{c} 
\newcommand{\AC}{\bar{\mathcal{A}}} 
\renewcommand{\AA}{\mathcal{A}}

\newcommand{\bp}{\bar{p}}

\usepackage{subcaption}

\begin{document}
\maketitle
\thispagestyle{empty}
\newpage
\tableofcontents
\section{Introduction}
In the past decades progress in the field of scattering amplitudes has improved our ability to make predictions in scattering experiments.\footnote{See Refs.
\cite{Britto:2010xq, Ellis:2011cr, Elvang:2013cua, Weinzierl:2016bus, Cheung:2017pzi} for reviews on the modern approach to scattering amplitudes and Ref.\cite{Heinrich:2020ybq} for 
a review of the state of the art of multi-loop calculations.} Recently,  techniques developed originally to tackle problems in perturbative quantum field theory are 
being applied to the study of classical observables and have led to insights about the classical limit, see e.g., 
\cite{Foffa:2016rgu, Cachazo:2017jef,  Cheung:2018wkq, Kosower:2018adc,Bjerrum-Bohr:2018xdl, Guevara:2018wpp, Blumlein:2019zku, Cristofoli:2019neg,
Maybee:2019jus, Arkani-Hamed:2019ymq, Bern:2019nnu, Bern:2019crd,Damgaard:2019lfh, Brandhuber:2019qpg, Bern:2020buy, Aoude:2020onz, Mogull:2020sak}. We would like to
single out two developments in scattering amplitudes. First, a generalisation of Feynman's tree theorem \cite{Feynman:1963ax} known as loop-tree duality
\cite{Catani:2008xa, CaronHuot:2010zt, Bierenbaum:2010cy, Bierenbaum:2012th, Buchta:2014dfa, Hernandez-Pinto:2015ysa, Runkel:2019yrs, Runkel:2019zbm, 
Aguilera-Verdugo:2020kzc, Verdugo:2020kzh, JesusAguilera-Verdugo:2020fsn}, which relates multi-loop amplitudes and phase space integrals and whose integrand is a tree-level 
amplitude-like object.\footnote{A non-exhaustive sample references on the relation between loops and trees is \cite{Brandhuber:2005kd, Huang:2015cwh, 
Feng:2016msc, Cachazo:2015aol,Geyer:2015bja,He:2016mzd,Ochirov:2017jby}.} The loop tree-duality can also be formulated in terms of response functions which can be obtained from the zero temperature Schwinger-Keldysh formalism 
\cite{Brandt:2006aj, CaronHuot:2010zt}. Second, the novel framework introduced by Kosower-Maybee-O'Connell (KMOC) \cite{Kosower:2018adc} to study classical observables from 
amplitudes. The KMOC formalism is based on expectation values of  operators between initial states and thus is akin to the zero temperature Schwinger-Keldysh 
formalism itself.

In this paper we are interested in relating these two ideas to thermal field theory. In their form however we cannot directly apply them since in thermal field theory one
is usually interested in currents (rather than scattering amplitudes) from which the thermodynamic properties of the system can be derived. Although the methods we will
develop can be applied to QED, QCD and gravity,  our primary example is a non-abelian plasma in the high temperature limit. 

The limit of high temperature $T$ in perturbative thermal QCD is useful for the description of collective phenomena in plasmas. This limit can be  consistently 
incorporated into an effective theory known as Hard Thermal Loop (HTL) effective theory~\cite{Braaten:1989mz,  Frenkel:1989br, Braaten:1990az, Taylor:1990ia,
Frenkel:1991ts}.  At the core of HTL effective theory is the resummation of 1-loop diagrams with the property of having external soft momenta and internal
hard loop momenta. Hard Thermal Loops (HTLs) are non-local currents which are not only gauge invariant but obey simple  Ward identities. Assuming 
a small coupling $g$ the soft  momenta are of order $gT$ and the hard momenta of order $T$. HTLs can be computed from the forward scattering of  thermal particles \cite{Frenkel:1991ts} which is reminiscent of the loop-tree duality for scattering
 amplitudes at zero temperature \cite{Catani:2008xa, CaronHuot:2010zt}. 
On the other hand,  HTLs can be reformulated in the language of classical  kinetic theory 
\cite{Blaizot:1993be, Blaizot:1994nr} and can be obtained from solutions of kinetic equations \cite{Kelly:1994ig}. 

The classicality of HTLs and their equivalence with solutions of kinetic equations raises the question of whether we can directly obtain them as a classical limit understood as
  the limit of $\hbar \rightarrow 0$ of a quantum current.  Inspired by the KMOC approach to classical observables,  we propose a map between classical limits of off-shell currents at zero temperature and HTL currents. This map is based on the simple
observation that to extract classical limits of scattering amplitudes one should distinguish  between the momentum $p$ of a particle and its wavenumber $\bp$ 
\begin{align}
 p \equiv \hbar \bar{p}, 
\end{align}
which we interpret as the distinction between soft and hard in HTL currents. As we will see, taking the classical limit will correspond to the high temperature 
limit. This scaling is the same one requires to study the classical dynamics of  gravitational waves from quantum field theory. Hence we will mostly inspired in the KMOC  approach but other strategies to obtain the classical limit may be used as well. e.g., 
\cite{Luna:2017dtq, Bjerrum-Bohr:2018xdl, Damgaard:2019lfh}. 

This paper is organised
as follows. In Section \ref{kin-current} we discuss briefly the equivalence of HTLs and solutions of kinetic equations and introduce the map to classical limits
of off-shell currents. In Section  \ref{QED-case} we consider QED. In Section \ref{QCD-gravity} we consider QCD and gravity. We discuss our
results in Section \ref{discussion}.

\section{From kinetic theory to classical limits of off-shell currents}
\label{kin-current}

The dynamics of non-abelian plasmas is described by QCD at finite temperature at distances  $\hbar/T \lesssim d \lesssim \hbar/(gT)$. For the rest of the paper we will use units  in which $k_B=c=1$ but
keeping $\hbar\ne 1$ since we are interested in  the study of the classical limit. The leading order thermal effects can be studied employing relativistic kinetic theory, which
is applicable at scales where the average distance between particles is of order $\hbar/T$ and such that the plasma is characterised by the Debye
length $\hbar/(gT)$.  
In the microscopic approach to kinetic theory\footnote{This approach has  been reviewed in \cite{Litim:2001db, Blaizot:2001nr}, see also Ref.\cite{Mrowczynski:2016etf} for
a recent review.},  we are interested in an ensemble of point-like particles characterised by a phase-space  distribution $f \equiv f(x, k, \cc)$, where the colour degrees
of freedom $c$ of the  plasma are treated as continuous classical variables. The phase space variables obey Wong equations~\cite{Wong:1970fu}
\begin{align}
	\frac{\dd k^\mu }{\dd \tau} =& g\, \cc^a(\tau)\, F^{a\,\mu\nu}\!(x(\tau))\,  v_{ \nu}(\tau)\,,  \label{Wong-momentum}
	\\
	\frac{\dd \cc^a}{\dd \tau}=& g f^{abc} v^\mu(\tau) A_\mu^b(x(\tau))\,\cc^c(\tau)\,,
	\label{Wong-color}
	\\
	D^\mu F_{\mu\nu}^a(x)=&  J^a_\nu (x)= g \int \dd \Phi(k) \dd \cc \; k_\nu \cc^a f(x, k, \cc),\,
	\label{YangMillsEOM}
\end{align}
where the Lorentz invariant phase space measure is defined by 
  \begin{align}
 \dd \Phi (k) \equiv \frac{\dd^4 k}{(2\pi)^4}  \Theta(k_0)\, 2\pi \delta (k^2-m^2).
\end{align}
Specialising to the case of $SU(3)$ the colour measure is defined by  
\begin{align}
\dd \cc \equiv \dd^8 \cc \;  \cc_R \delta(\cc^a \cc^a-q_2)\delta(d_{abc}\cc^a\cc^b\cc^c-q_3),    
\end{align}
where the colour part ensures the conservation of the Casimir invariants $q_i$  and $c_R$ is a normalisation constant defined such that $ \int \dd \cc=1$\footnote{Notice that
the measure is defined in terms of the dimensionless constant $c^a/\hbar$.}. In the collision-less case the 
Vlasov-type equation for  the distribution function $f(x, k, \cc)$ is given by
\begin{align}
&\frac{\dd f }{\dd \tau}=k^\mu \left( \frac{\partial }{\partial x^\mu} -g f^{abc} A_\mu^b \cc^c \frac{\partial }{\partial \cc^a} -g \cc^a F^a_{\mu\nu} 
\frac{\partial}{\partial k_\nu}\right) f=0. 
\label{Vlasov} 
\end{align}
It is well-known that both Wong and Vlasov equations can be derived from suitable classical limits \cite{Brown:1979bv}. The Vlasov equations arises from the classical limit of 
the corresponding equation for the Wigner operator \cite{Heinz:1983nx, Heinz:1984yq, Elze:1989un}.

In this paper however we are interested in the classical limits of currents leading to \emph{solutions} of the kinetic equations rather than a derivation of kinetic 
equations themselves.  Eq.\eqref{Vlasov} can be solved iteratively by  expanding  $f(x, k, \cc)$ around the equilibrium state. In equilibrium the distribution function 
only depends on the  energy $k_0$ of the system and we can  expand it in powers of the coupling $g$
\begin{align}
 f(x, k, \cc)= f^{(0)}(k_0)+ \Delta^{(1)} f (x,k,c) +  \Delta^{(2)} f (x,k,c) + \dots, 
\end{align}
where $\Delta^{(i)} f (x,k,c)$ is of order $\mathcal{O}(g^i)$. We can then use the equilibrium distribution function to solve for $\Delta^{(1)} f(x,k,c)$ and
re-insert it to find $\Delta^{(2)} f (x,k,c)$ and so on. From Eq.\eqref{YangMillsEOM}  we can write the total current 
as\footnote{The term of order $\mathcal{O}(g^0)$ vanishes due to the
identity 
$\int \dd \cc \ \cc^a=0$.}
\begin{align}
 J_a^\mu (x)=  \Delta^{(1)} J_a^\mu (x)+ \Delta^{(2)} J_a^\mu (x)+\dots.
 \label{sols-current-Vlasov}
\end{align}
The currents can be expressed in the form 
\begin{align}
J_\mu^a(x)= \Pi^{ab}_{\mu\nu} A_b^\nu+ \frac{1}{2}\Pi^{abc}_{\mu\nu\rho} A_b^\nu A_c^\rho+ \dots, 
\label{HTL-expansion}
\end{align}
where $\Pi^{ab \cdots}_{\mu\nu\cdots}$ are thermal currents\footnote{The current has to be understood as a functional of the soft gauge fields.}, which can be matched
against those  obtained from the high temperature limit of thermal QCD (See e.g., Ref.\cite{Laine:2016hma}).
One can thus conclude that  HTL are classical \cite{Kelly:1994ig, Kelly:1994dh}. Notice that despite our choice of units we will keep our classical quantities 
independent of $\hbar$ as they should.  The setup leading to Eq.\eqref{sols-current-Vlasov} is reminiscent of the  one to obtain perturbative solutions from 
classical equations of motion \cite{Goldberger:2016iau, Damour:2017zjx, Goldberger:2017frp, 
Goldberger:2017vcg,Goldberger:2017ogt,  Luna:2017dtq,  Chester:2017vcz, Shen:2018ebu} which can be matched to classical limits of scattering amplitudes. Within 
the KMOC formalism this is done by defining certain observables which are well-defined both classically and quantum mechanically~\cite{Kosower:2018adc, Maybee:2019jus, delaCruz:2020bbn}. Besides the distinction between the momentum $p$ of a particle and its wave-number
$\bp$ the introduction of appropriate coherent states is required to take the classical limit. Since we are interested in off-shell currents rather than observables,
we cannot  directly apply the KMOC formalism but as we will see many of its characteristics remain.  

The classical limit of amplitudes in the KMOC formalism is slightly different in QED and QCD due to the presence of colour in the latter. In QCD 
the dimensionless coupling $g$ scales as $\bar{g} \sqrt{\hbar}$.  In QED the dimensionless coupling $e$ scales as $\bar{e}/\sqrt{\hbar}$ and similarly for
gravity.\footnote{See Ref.\cite{delaCruz:2020bbn} for a longer discussion about this point.} The
scaling in QCD is complemented by scaling of the colour factors so ultimately we can use the same procedure  in Ref.\cite{Kosower:2018adc} to obtain the 
classical limit. Adopting  the conventions of Ref.\cite{delaCruz:2020bbn} we have
\begin{align}
 [\mathbb{C}^a, \mathbb{C}^b]=\ii \hbar f^{abc}\mathbb{C}^c,
 \label{lie-algebra}
\end{align}
emphasising that $\mathbb{C}$ corresponds to an operator and
\begin{align}
 \braket{p_i |  \mathbb{C}^a |p^j}\equiv (C^a)_{i}^{\ j}= \hbar (T^a)_i^{\ j}.
\end{align}
The corresponding classical colour charge $c^a\equiv \braket{\psi|\mathbb{C}^a|\psi}$ is  obtained from appropriate  coherent states $\ket{\psi}$. In the spinless case
gravity does not bring any new ingredient other than complexity so the KMOC algorithm follows the QED one. 

The interplay between soft and hard momenta is very generally linked to the classical limit, usually through the Eikonal approximation, 
and thus it is tempting to relate it the   HTL approximation. 
 However, \emph{operationally} in the  HTL approximation  one integrates over  hard momenta running in the loop $k \sim T$  and considers 
 external soft momenta $p \sim g T$. The results are then expanded in powers of $|p|/|k|$,  see Ref.\cite{Laine:2016hma} for a summary of the 
rules to compute HTLs. 
By associating classical wavenumbers to the external soft momenta such that they scale with $\hbar$ an expansion in powers of 
$\hbar$ will correspond to an expansion in high temperature. This formal equivalence can also be observed at the level of the dimensional reduced 
effective action.\footnote{See e.g., Chapter 38 in Ref.\cite{ZinnJustin:1989mi}.} Physically this has the interpretation of  matching the high temperature regime to the
classical regime. 	

In momentum space the coefficients in Eq.\eqref{HTL-expansion} can be written as follows \cite{Barton:1989fk, Frenkel:1991ts}
\begin{align}
\Pi^{a_1 a_2 \dots a_n}_{\mu_1 \mu_2 \cdots \mu_n} (p)= 
\int \dd \Phi (k) N(k_0) \mathcal{H}^{a_1 a_2 \dots a_n}_{\mu_1 \mu_2 \cdots \mu_n}(k, p), \label{forward}
\end{align}
where $N(k_0)$ is some distribution function and $\mathcal{H}_n\equiv \mathcal{H}^{a_1 a_2 \dots a_n}_{\mu_1 \mu_2 \cdots \mu_n}(k, p)$ is the integrand of the HTL current. If we are 
interested in fermions we will define the distribution function with a minus sign due to the presence of a fermion loop. We have used the short-hand notation $p$ to express the dependence on the full set of external momenta $p_1, \dots, p_n$. A generating functional has been given in 
\cite{Taylor:1990ia}.\footnote{See Ref.\cite{Elmfors:1998wd} for a simple derivation of the 
functional and Refs.\cite{Efraty:1992gk, Efraty:1992pd, Jackiw:1993zr, Nair:1993rx, Nair:1994xs} for the relation of the generating functional
and  Chern-Simons theory.}

\begin{figure}[ht]	
\centering
\begin{tikzpicture}[thick, transform shape]
\begin{feynman}
\vertex (c0);
\vertex [below =1 of c0](c1);
\vertex [left =1 of c1](c2);
\vertex [left =2 of c1](c3);
\vertex [right =1 of c1](c4);
\vertex [right =2 of c1](c5);
\vertex[below = 0.3 of c0](d1){\( \dots \)};
\vertex [left =1 of c0](p2);
\vertex [left =2 of c0](p3);
\vertex [above =0.2 of p3](p3up){\(p_1\)};
\vertex [above =0.2 of p2](p2up){\(p_2\)};
\vertex [right =1 of c0](p4);
\vertex [above =0.2 of p4](p4up){\(p_{n-1}\)};
\vertex [right =2 of c0](p5);
\vertex [above =0.2 of p5](p5up){\(p_n\)};
\vertex [right = 3 of c1] (l1) {$k$};
\vertex [left = 3 of c1] (l2) {$k$};
\diagram* {
(c2) -- [boson, edge label=] (p2),
(c3) -- [boson, edge label=] (p3),
(c4) -- [boson, edge label=] (p4),
(c5) -- [boson, edge label=] (p5)
};
\end{feynman}
\begin{scope}[decoration={
		markings,
		mark=at position 0.6 with {\arrow[>=Stealth]{>}}}] 
\draw[postaction={decorate}] (l2) -- node [right=4pt] {}(c3);
\draw[postaction={decorate}] (c3) -- node [right=4pt] {}(c2);
\draw[postaction={decorate}] (c2) -- node [right=4pt] {}(c4);
\draw[postaction={decorate}] (c4) -- node [right=4pt] {}(c5);
\draw[postaction={decorate}] (c5) -- node [right=4pt] {}(l1);
\end{scope}
\end{tikzpicture}	
\caption{Typical comb diagrams required for the calculation of currents in the high temperature limit. The wavy lines represent gauge bosons. 
The solid lines represent massive particles such as scalars or quarks, or massless particles such as gauge bosons or ghosts.}
\label{comb-diagrams}
\end{figure}
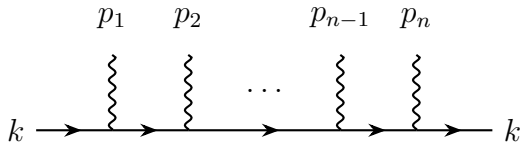

\begin{figure}
\centering
 \begin{tikzpicture}[thick, transform shape]
\begin{feynman}
\vertex(ce);
\vertex [above = 1 of ce] (c0);
\vertex [left = 1 of ce] (c1){\(k\)};
\vertex [right = 1 of ce] (c2){\(k\)};
\vertex [above left = 1 of ce] (c3){\(p_1, \mu_1, a_1\)};
\vertex [above right = 1 of ce] (c4){\(p_n, \mu_n, a_n\)};
\vertex [above = 0.5 of ce] (c5){\(\dots\)};
\diagram* {
c0[blob]--(c1),
c0[blob] --(c2),
c0[blob] - - [boson, edge label'=\(\)](c3),
c0[blob]- - [boson, edge label'=\(\)](c4),
};
\begin{scope}[decoration={
		markings,
		mark=at position 0.6 with {\arrow[>=Stealth]{>}}}] 
\draw[postaction={decorate}] (c1) -- node [right=4pt] {}(c0);
\draw[postaction={decorate}] (c0) -- node [right=4pt] {}(c2);
\end{scope}
\end{feynman}
\end{tikzpicture}
\caption{Off-shell current. The blob represents a sum over tree-level Feynman diagrams. Off-shell gauge bosons can be photons, gluons or
gravitons.}
\label{current-off-shell}
\end{figure}
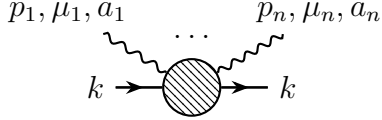

In thermal perturbation theory the integrands  in Eq.\eqref{forward} can be constructed by considering permutations of comb diagrams (See Fig.\ref{comb-diagrams}). This approach can be applied systematically to a variety of 
theories including gravity \cite{Brandt:1992dk, Brandt:1993bj, Brandt:1997se, Brandt:1998gb, Brandt:1999gf, Brandt:2002aa, Brandt:2002qk, Brandt:2006aj,Brandt:2015fna}.  
For on-shell amplitudes at zero temperature Eq.\eqref{forward} is the usual forward limit \cite{Brandt:2006aj, Catani:2008xa, CaronHuot:2010zt} and hence a 
current at zero temperature is a natural  candidate to relate the classical limit and the high temperature limit. Therefore we
consider the current represented in Fig.\ref{current-off-shell} which can be computed using the Feynman rules of the theory under consideration. 
These currents have the property of having $n$-particles off-shell while the momentum of the massive particle running in the loop is on-shell, that is
$k^2=m^2$. 

The forward limit is in general singular therefore we must consider a regularisation scheme.  We define $\mathcal{F}$ as the set of all Feynman graphs  and 
$\mathcal{S}$ as the set of tadpole graphs --- i.e. those graphs containing a zero-momentum internal edge (see Fig.\ref{4pt-diags}). Suppressing colour and Lorentz 
indices, we consider the following \emph{regularised} current
\begin{align}
 \AA_n(k, p_1, \cdots, p_n, k) \equiv \sum\limits_{G \in  \mathcal{F} \setminus \mathcal{S}} f(G),
\end{align}
where $f(G)$ is a rational expression of the form $N(G)/D(G)$. The regularised current is simply obtained by removing tadpole graphs.\footnote{See 
Ref.\cite{Runkel:2019zbm} for a general version of this idea within loop-tree duality.} In the following we assume that the current is regulated. The 
classical limit of this current will be defined by
\begin{align}
 \AC_n(k, \bp_1, \cdots, \bp_{n}) \equiv  \widetilde{\text{Tr}} \left(\lim_{\hbar \rightarrow 0} \AA_n(k, \hbar \bp_1, \cdots, \hbar \bp_{n}, k) \right),
 \label{current-general}
\end{align}
which is obtained by performing a Laurent expansion in powers of $\hbar$ after rescaling the momenta of the soft particles and couplings following the algorithms in 
Refs.~\cite{Kosower:2018adc, Maybee:2019jus, delaCruz:2020bbn}. The operator $\widetilde{\text{Tr}}$ depends on whether the theory is coloured or not. We define it by 
\begin{align}
\widetilde{\text{Tr}} (\bullet) \equiv \begin{cases} 	
          \hbar^{n-2} \text{Tr} (\bullet) & \text{QCD}, \\
          \text{Id} (\bullet) & \text{QED} \ \text{and} \ \text{gravity},
       \end{cases}
              \label{trace-tilde-def}
\end{align}
where $\text{Id}$ is the identity operator and the $\hbar^{n-2}$ is required on dimensional grounds. We adopt the convention that our classical results will depend on the 
dimensionless coupling $g=\bar{g} \sqrt{\hbar}$ and $e=\bar{e} /\sqrt{\hbar}$ and that the external momenta is associated to wavenumbers. If desired one may restore the 
dependence on momenta by dimensional analysis. In the case of QCD, this requires an overall inverse power of $\hbar$ in Eq.\eqref{trace-tilde-def}. 
The currents just defined are in general gauge dependent. Similar currents occur for example in the Berends-Giele recursion
\cite{Berends:1987me}.

We propose to map the integrand in \eqref{forward} in momentum space to the classical limit of the $n$-point forward current 	
\begin{align}
   \AC_n(k, \bp_1, \cdots, \bp_{n}) \leftrightarrow \mathcal{H}_n(k, p_1, \dots, p_n),
\end{align}
and similarly for QED and gravity.

Using the imaginary-time formalism, where time is traded with the temperature and the structure of the propagator is similar to the zero temperature case, we can relate zero-temperature amplitudes and finite-temperature amplitudes by 
analytic continuation, see e.g.,\cite{Das:1997gg}. The currents we are proposing are amplitude-like objects and their computation is not based on sums over permutations of comb diagrams 
(see Fig.\ref{comb-diagrams}) as the usual forward scattering  approach to thermal field theory \cite{Frenkel:1991ts}. Let us also remark that although we are considering the case where only massive particles appear in the loop, this case can also be
utilised to  study the case in which hard gauge particles run in the loop since the tensor structures are the same.\footnote{We thank Fernando T. Brandt for this 
insight.} Thus it is enough to discuss the case with matter loops. The pure gauge calculation with ghosts included appears e.g., in Ref. \cite{Brandt:2006aj}. It is in 
this sense that ghosts will not be required. From an amplitudes point of view this is a simple consequence of our calculation being a tree-level one.

In order to model the non-abelian plasma we use the lagrangian  
\begin{equation}
\mathcal{L} = \left[ (D_\mu \varphi^\dagger) D^\mu \varphi - \frac{m^2}{\hbar^2} \varphi^\dagger 
\varphi\right] - \frac14 F^a_{\mu\nu} F^{a\,\mu\nu}, \label{eqn:lagrangian}
\end{equation}
where  $D_\mu = \partial_\mu + i g A_\mu^a T^a$ and the field strength tensor  $F^a_{\mu\nu}=\partial_\mu A^a_\nu-\partial_\nu A^a_\mu 
-g f^{abc} A_\mu^b A_\nu^c$. For QED we employ the abelian version of this lagrangian.  Let us remark that Eq.\eqref{Vlasov} and its abelian version
describe a plasma for \emph{spinless} particles  and therefore this lagrangian is the appropriate to consider in the classical limit\footnote{
See Ref.\cite{Kraemmer:1994az} for the use of scalar QED as a toy model for gluonic QCD.}. We use Feynman gauge for computations and adopt the following 
normalisation of the generators of the Lie algebra  
\begin{align}
 \text{Tr}\left( C^a C^b\right)= \frac{\hbar^2}{2} \delta^{ab}.
\label{trace-def}
\end{align}
\section{QED}
\label{QED-case}
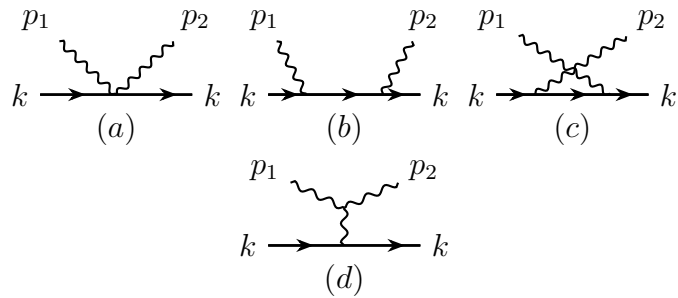
\begin{figure}[ht]
\centering
\begin{tikzpicture}[thick, transform shape]
\begin{feynman}
\vertex(ce);
\vertex [below = 0.1 of ce] (ce01){\((b)\)};
\vertex [left = 0.5 of ce] (l01);
\vertex [right = 0.5 of ce] (r01);
\vertex [above = 1 of ce] (c02);
\vertex [left = 1 of ce] (c12){\(k\)};
\vertex [right = 1 of ce] (c22){\(k\)};
\vertex [above left = 1 of ce] (c32){\(p_1\)};
\vertex [above right = 1 of ce] (c42){\(p_2\)};

\vertex[left = 3 of ce](ce2);
\vertex [below = 0.1 of ce2] (ce02){\((a)\)};
\vertex [above = 1 of ce2] (c01);
\vertex [left = 1 of ce2] (c11){\(k\)};
\vertex [right = 1 of ce2] (c21){\(k\)};
\vertex [above left = 1 of ce2] (c31){\(p_1\)};
\vertex [above right = 1 of ce2] (c41){\(p_2\)};

\vertex[right = 3 of ce](ce3);
\vertex [below = 0.1 of ce3] (ce03){\((c)\)};
\vertex [left = 0.5 of ce3] (l02);
\vertex [right = 0.5 of ce3] (r02);

\vertex [above = 1 of ce3] (c03);
\vertex [left = 1 of ce3] (c13){\(k\)};
\vertex [right = 1 of ce3] (c23){\(k\)};
\vertex [above left = 1 of ce3] (c33){\(p_1\)};
\vertex [above right = 1 of ce3] (c43){\(p_2\)};

\vertex[below = 2 of ce](ce4);
\vertex [below = 0.1 of ce4] (ce04){\((d)\)};
\vertex[above = 0.5 of ce4](ce5);

\vertex [above = 1 of ce4] (c04);
\vertex [left = 1 of ce4] (c14){\(k\)};
\vertex [right = 1 of ce4] (c24){\(k\)};
\vertex [above left = 1 of ce4] (c34){\(p_1\)};
\vertex [above right = 1 of ce4] (c44){\(p_2\)};

\diagram* {
(c12)--(c22),
(c11) --(c21),
(c13) --(c23),
(c14) --(c24),
(ce2) - - [boson, edge label'=\(\)](c31),
(ce2) - - [boson, edge label'=\(\)](c41),
(l01)- - [boson, edge label'=\(\)](c32),
(r01)- - [boson, edge label'=\(\)](c42),
(l02)- - [boson, edge label'=\(\)](c43),
(r02)- - [boson, edge label'=\(\)](c33),
(ce5)- - [boson, edge label'=\(\)](ce4),
(ce5)- - [boson, edge label'=\(\)](c34),
(ce5)- - [boson, edge label'=\(\)](c44),
};
\begin{scope}[decoration={
		markings,
		mark=at position 0.6 with {\arrow[>=Stealth]{>}}}] 
\draw[postaction={decorate}] (c12) -- node [right=4pt] {}(c22);
\draw[postaction={decorate}] (c13) -- node [right=4pt] {}(c23);
\end{scope}
\begin{scope}[decoration={
		markings,
		mark=at position 0.3 with {\arrow[>=Stealth]{>}}}] 
\draw[postaction={decorate}] (c11) -- node [right=4pt] {}(c21);
\draw[postaction={decorate}] (c14) -- node [right=4pt] {}(c24);
\end{scope}
\begin{scope}[decoration={
		markings,
		mark=at position 0.9 with {\arrow[>=Stealth]{>}}}] 
\draw[postaction={decorate}] (c11) -- node [right=4pt] {}(c21);
\draw[postaction={decorate}] (c14) -- node [right=4pt] {}(c24);
\end{scope}
\begin{scope}[decoration={
		markings,
		mark=at position 0.2 with {\arrow[>=Stealth]{>}}}] 
\draw[postaction={decorate}] (c12) -- node [right=4pt] {}(c22);
\draw[postaction={decorate}] (c13) -- node [right=4pt] {}(c23);
\end{scope}
\begin{scope}[decoration={
		markings,
		mark=at position 0.9 with {\arrow[>=Stealth]{>}}}] 
\draw[postaction={decorate}] (c12) -- node [right=4pt] {}(c22);
\draw[postaction={decorate}] (c13) -- node [right=4pt] {}(c23);
\end{scope}
\end{feynman}
\end{tikzpicture}
\caption{Diagrams related to the 2-point function in QED, QCD and gravity: Diagrams (a)-(c) contribute to the regularised forward limit. Diagram (d) leads to 
a 1-loop diagram with zero momentum, i.e.,  a tadpole.}
\label{4pt-diags}
\end{figure}

As our first example we will consider a QED plasma where the equilibrium distribution is given by the Fermi-Dirac distribution
\begin{align}
 N(k_0)\equiv-\frac{1}{\exp(k_0/T)+1},
\end{align}
where the sign is included to indicate that we are interested in matching the high temperature limit of $n$-point functions in QED with a fermion running in the 
loop. Recall that in our approach however the classical limit is taken from scalar QED since we are solving the spinless Vlasov equation. The regulated current is simply
\begin{align}
\AA_{n}(k, \hbar \bp_1, \cdots, \hbar \bp_{n}, k),
\end{align}
where all photons are outgoing. Let us consider the simple case $n=2$ whose contributing diagrams are
shown in Fig.\ref{4pt-diags}. Using the rescaled momenta a simple calculation leads to 
\begin{align}
 &\AA^{\mu\nu}(k, \hbar \bp_1, \hbar \bp_{2}, k)=2 e^2 \Big( \eta^{\mu \nu }+ 
\frac{k^{\nu } \left(4 \bar{p}_1{}^2 k^{\mu }-4 k\cdot \bar{p}_1 \bar{p}_1{}^{\mu }\right)
-4 k^{\mu } k\cdot \bar{p}_1 \bar{p}_1{}^{\nu }+\hbar ^2 \bar{p}_1{}^2 \bar{p}_1{}^{\mu } \bar{p}_1{}^{\nu }}{4 \left(k\cdot \bar{p}_1\right){}^2
-\hbar ^2 \left(\bar{p}_1{}^2\right){}^2}
\Big), \nonumber
\end{align} 
where we have used  $p_2=-p_1$. Upon performing a Laurent expansion in powers of $\hbar$ we obtain 
\begin{align}
\AC^{\mu\nu}(k,\bp)= 2 e^2 
 \left(\eta^{\mu \nu }+\frac{\bar{p}_1{}^2 k^{\mu } k^{\nu }}{\left(k\cdot \bar{p}_1\right){}^2}-\frac{k^{\nu} \bar{p}_1{}^{\mu}
 +k^{\mu } \bar{p}_1{}^{\nu }}{k\cdot \bar{p}_1}\right),
\end{align}
which recovers the $\mathcal{O}(T^2)$ HTL 2-photon integrand.

Notice that the propagators considered for our currents do not have an 
$\ii \epsilon$ term. In order to recover the retarded temperature dependent currents we consider the analytic continuation $\bp^0_n \rightarrow 
\bp^0_n +\ii \epsilon$
and $\bp^0_i \rightarrow \bp^0_i -\ii \epsilon $, for $i=1, \dots, n-1$,
where we assume that the vertex corresponding to $\bp_n$ corresponds to the one with the largest time~\cite{Brandt:1998gb}. Finally, we can perform the radial integration
by parametrising $k^\mu=|\mathbf{k}| K^\mu$ with $K^\mu =(\sqrt{1+m^2/|\mathbf{k}|^2}, \mathbf{k}/|\mathbf{k}|)$~\cite{Brandt:1992dk, Brandt:2006aj, Brandt:2015fna} (See 
Appendix \ref{Atemp}). 

This simple example summarises our approach in general: first consider the $n$-point regulated current \eqref{current-general},  then compute the current using 
Feynman graphs or Berends-Giele recursions, etc, and calculate the classical limit with the replacement  $\hbar \bp_i$ for the gauge bosons. Physical propagators are obtained by 
analytic continuation.

In Ref.\cite{Frenkel:1991ts}  ``super-leading'' terms in the temperature expansion appear, which cancel after all permutations of graphs are considered. In 
our approach we are not considering permutations (e.g., in the $n=2$ case we are only summing  3.a-3.c in Fig.\ref{4pt-diags}) so it is interesting to ask if 
``superleading'' terms appear here too. In our approach these terms have the form $k^{\mu_1} \cdots k^{\mu_n}/(\hbar^{n-1}(k \cdot \bp_1) \cdots
(k \cdot \bp_{n-1}))$ and hence they correspond to (if any) singular terms in the classical limit.
In our general expression in Eq.\eqref{current-general} these terms cancel (QED) or vanish (QCD). The HTL currents truly arise as classical limits of the currents 
$\AA^{\mu\nu}(k, \hbar \bp_1, \hbar \bp_{2}, k)$. For a neutral QED plasma it  is well-known that higher point functions are either vanishing or 
subleading, see e.g.,  \cite{Bellac:2011kqa}. We can however consider a non-neutral plasma which would still be described by the Vlasov equation and thus 
study generalisations of HTL for higher point functions in QED\cite{Brandt:2015fna}. 
After taking the classical limit and keeping terms of $\mathcal{O}(\hbar^0)$ it is straightforward to compute higher order corrections in the 
temperature. We have checked up to $n=4$ that our methods reproduce those obtained from the thermal field theory approach\cite{Brandt:2015fna}. 
A 3-point example is given in  Appendix~\ref{QED-3pt}.

\section{QCD and gravity}
\label{QCD-gravity}
The QED case illustrates the procedure of taking the classical limit for the kinematics. The new ingredient in QCD is colour. 
The lowest order  contribution requires the calculation of a diagram with a three-gluon vertex as shown in Fig.\ref{4pt-diags}, which leads to a tadpole
diagram and therefore removed.\footnote{The contribution from three-gluon vertexis typically neglected because it does not correspond to a one-particle 
irreducible graph.}
Notice that the first diagram in Fig.\ref{4pt-diags}  does not lead to a tadpole since the internal  edge with zero momentum is not present. Computing the current and
taking the classical limit leads to      
\begin{align}
&\AA^{\mu_1\mu_2}_{a_1 a_2}(k, \hbar \bp_1, \hbar \bp_2)=\\
&\qquad \qquad \frac{\bar{g}^2}{\hbar} \Bigg[
 C^{a_1}\cdot C^{a_2} \left(-\frac{2 k^{\mu_1 } k^{\mu_2 }}{\hbar k\cdot \bar{p}_1}+\frac{\bar{p}_1{}^2 k^{\mu_1 } k^{\mu_2 }}{\left(k\cdot \bar{p}_1\right){}^2}
 -\frac{k^{\nu } \bar{p}_1{}^{\mu }}{k\cdot \bar{p}_1}-\frac{k^{\mu_1 } \bar{p}_1^{\mu_2 }}{k\cdot \bar{p}_1}+\eta^{\mu_1 \mu_2 } + \mathcal{O}(\hbar)\right) \nonumber\\
 &\qquad \qquad \quad +C^{a_2}\cdot C^{a_1} \left( 
\frac{2 k^{\mu_1 } k^{\mu_2 }}{\hbar k\cdot \bar{p}_1}+\frac{\bar{p}_1{}^2 k^{\mu_1 } k^{\mu_2 }}{\left(k\cdot \bar{p}_1\right){}^2}-\frac{k^{\mu_2 }
\bar{p}_1{}^{\mu_1 }}{k\cdot \bar{p}_1}-\frac{k^{\mu_1 } \bar{p}_1{}^{\mu_2 }}{k\cdot \bar{p}_1}+\eta^{\mu_1 \mu_2 } + \mathcal{O}(\hbar)\right)
 \Bigg],\nonumber
\end{align} 
where $C^a\cdot C^b \equiv {(C^a)}^{i}_{\ j} {(C^b)}^{j}_{ \ k}$. At first sight this expression contains singular terms of the form $k^{\mu_1} k^{\mu_2}/(\hbar k\cdot \bar{p}_1)$, which we
mentioned previously. However using the Lie
algebra \eqref{lie-algebra} these produce classical terms  with a vanishing trace. Therefore after tracing  we obtain the well-known result for QCD
\begin{align}
&\AC^{\mu_1\mu_2}_{a_1 a_2}(k, \bp)=  \delta^{a_1 a_2 } \Pi^{\mu_1\mu_2}(\bp_1)  \equiv \delta^{a_1 a_2 } g^2 \left(\frac{\bar{p}_1{}^2 k^{\mu_1 }
k^{\mu_2 }}{\left(k\cdot \bar{p}_1\right){}^2}
 -\frac{k^{\mu_2 } \bar{p}_1{}^{\mu_1 }}{k\cdot \bar{p}_1}-\frac{k^{\mu_1 } \bar{p}_1{}^{\mu_2 }}{k\cdot \bar{p}_1}+\eta^{\mu_1 \mu_2 }\right). 
 \end{align}
Notice that the final result is expressed in terms of the dimensionless coupling $\bar{g}\sqrt{\hbar}$. We now consider the three-point function. We decompose the current in a basis of colour factors as follows
\begin{align}
&\AA^{\mu_1\mu_2\mu_3}_{a_1 a_2 a_3}(k, \hbar \bp_1, \hbar \bp_{2}, \hbar \bp_{3}, k)=
\sum\limits_{\sigma \in S_3} C^{a_{\sigma_1}}\cdot C^{a_{\sigma_2}} \cdot C^{a_{\sigma_3}} J^{\mu_1\mu_2\mu_3}(\sigma_1, \sigma_2, \sigma_3).  \nonumber
\end{align}
The kinematic coefficients $J^{\mu_1\mu_2\mu_3}(\sigma_1, \sigma_2, \sigma_3)$  are straightforward to compute but lengthy and not presented here.
The leading contribution in the classical limit is of order $\mathcal{O}(1/\hbar)$, i.e.,   singular at first sight. However upon tracing we can bring the result into
the form
 \begin{align}
& \AC^{\mu_1\mu_2\mu_3}_{a_1 a_2 a_3}(k, \bp_1,  \bp_{2},\bp_{3}, k)= 2  \frac{\bar{g}^3 \hbar}	{\hbar^{3/2}} [\text{Tr} \left(C^{a_1}\cdot C^{a_3} \cdot C^{a_2}  
\right)-\text{Tr}
\left(C^{a_1}\cdot C^{a_2} \cdot C^{a_3}  \right)] \frac{A_{\text{QED}}^{\mu_1\mu_2\mu_3}}{\hbar}, \label{3-point-QCD}
\end{align}
where 
\begin{align}
 A_{\text{QED}}^{\mu_1\mu_2\mu_3}= \sum\limits_{ \sigma \in \text{Cyclic}}\Bigg[&
 \frac{2k^{\mu_{\sigma_1}} k^{\mu_{\sigma_2}}}{k\cdot \bp_{\sigma_3}} \left(\frac{\bp_{\sigma_1}^{\mu_{\sigma_3}}}{k\cdot \bp_{\sigma_1}}- 
\frac{\bp_{\sigma_2}^{\mu_{\sigma_3}}}{k\cdot \bp_{\sigma_2} } \right) 
 \label{QED-like-current}\\
&+k^{\mu_{ \sigma_1}} k^{\mu_{ \sigma_2}} k^{\mu_{ \sigma_3}} \left( \frac{\bp_{\sigma_1}^2}{ (k\cdot \bp_{\sigma_1})^2} 
\left(\frac{1}{k\cdot \bp_{\sigma_2}}-\frac{1}{k\cdot \bp_{\sigma _3}} \right)\right) \Bigg],\nonumber
\end{align}
where $\text{Cyclic}$ is the set of cyclic permutations of $\{1,2,3\}$. Notice that in the commutative case Eq.\eqref{3-point-QCD} is simple telling us that singular
terms vanish in the classical limit since the operator $\widetilde{\text{Tr}}$ is replaced by the identity operator.  Using Eqs.\eqref{lie-algebra} and \eqref{trace-def} we find 
\begin{align}
 \AC^{\mu_1\mu_2 \mu_3 }_{a_1 a_2 a_3}(k,  \bp) = \ii g^3 f^{a_1 a_2 a_3} A^{\mu_1\mu_2\mu_3}_{\text{QED}},
\end{align}
where the current satisfies the identity 
\begin{align}
\bp_{3 \mu_3} \AC^{\mu_1 \mu_2 \mu_3}_{a_1 a_2 a_3}(k,  \bp)= \ii g f^{a_1 a_2 a_3} 
\left[\Pi^{\mu_1 \mu_2 }(p_1)-\Pi^{\mu_1 \mu_2 }(p_2)\right].
\end{align}
Remarkably, we find that the kinematic structure of the current is encoded in the commutative QED part in a similar way as the classical colour and momentum impulse 
observables \cite{delaCruz:2020bbn}. Although we can continue computing the next contributions in a similar fashion, it is well-known we can reconstruct higher 
point functions using Ward 
identities and the above relation \cite{Frenkel:1989br, Braaten:1990az}. The expression \eqref{QED-like-current} agrees with previously computed expressions 
in Refs.\cite{Brandt:2002aa, Brandt:2002rw} 

There is no new ingredient in gravity regarding the classical limit and therefore we can apply our formalism for this case too. HTL for gravity, reviewed
in  Ref.\cite{Kraemmer:2003gd}, are relevant for the physics of the early universe. They have been investigated in Refs. \cite{Brandt:1993bj, Brandt:1998hd}. 
We expand in powers of the graviton coupling $\kappa$ with  $\kappa^2=32\pi G$ and set $g_{\mu_1\mu_2}=\eta_{\mu_1\mu_2}+\kappa h_{\mu_1\mu_2}$. The 2-point graviton  function 
can be computed from the diagrams in Fig.\ref{4pt-diags} with the gluon replaced by  a graviton, where as in QCD we regulate the current by removing contributions from 
the 3-graviton vertex.  Using the conventions in Ref.\cite{Holstein:2006bh} the calculation is straightforward and leads to 

\begin{align}
 \AC^{\alpha\beta; \gamma\delta}(k, \bp)= \frac{1}{2} \kappa^2 \Bigg[&\frac{\bar{p}_1{}^2 k^{\alpha } k^{\beta } k^{\gamma } 
k^{\delta }}{\left(k\cdot \bar{p}_1\right){}^2}-
\frac{k^{\beta } k^{\gamma } k^{\delta } \bar{p}_1{}^{\alpha }+k^{\alpha } k^{\gamma } k^{\delta } \bar{p}_1{}^{\beta }+k^{\alpha } k^{\beta}k^{\delta }
\bar{p}_1{}^{\gamma }-k^{\alpha } k^{\beta } k^{\gamma } \bar{p}_1{}^{\delta }}{k\cdot \bar{p}_1} \nonumber\\
&+\eta^{\beta \gamma } k^{\alpha } k^{\delta }+\eta^{\beta \delta } k^{\alpha } k^{\gamma }+\eta^{\alpha \gamma} 
k^{\beta } k^{\delta }+\eta^{\alpha \delta } k^{\beta } k^{\gamma }\Bigg],
\end{align}
which agrees with the result in Ref.\cite{Brandt:1993bj}. In general the 2-point graviton function depends on the representation of the graviton 
field but one can redefine it in terms of the above expression to obtain a 2-point function independent of the graviton parametrisation
\cite{Brandt:1993bj}.

\section{Discussion}
\label{discussion}
In this paper we have shown that HTLs arise from classical limits of off-shell currents.   The classical nature of  hard-thermal loop amplitudes was made manifest by relating the momenta of the 
soft particles to wavenumbers. The classical limit is then obtained following the KMOC algorithm. In this way, the 
high temperature limit is formally equivalent to an expansion in powers of $\hbar$ thus allowing a map  
between HTL amplitudes and classical limits of off-shell currents. The off-shell currents encode the information of permutation 
of comb diagrams and can be easily computed from Feynman diagrams or Berends-Giele recursions. Since our off-shell currents in the classical 
limit are gauge invariant and satisfy Ward identities their ``on-shell'' properties would be interesting to study, in particular the Britto-Cachazo-Feng-Witten 
\cite{Britto:2005fq} recursion, the colour-kinematics duality and the double copy\cite{Bern:2008qj, Bern:2010ue, Bern:2019prr}. For general off-shell currents 
these properties are generally more difficult to make manifest  than for amplitudes \cite{Britto:2012qi, Mastrolia:2015maa, Jurado:2017xut}.  At the classical level the double copy 
is much more flexible  but the idea of a formal replacement between colour and kinematics is preserved \cite{Monteiro:2014cda}. This classical double copy
is the appropriate to relate gravity and QCD in the high temperature limit.

On the more phenomenological side the inclusion of spin and collisions\footnote{Collision functions in the context of classical transport theory have been studied e.g., in \cite{Litim:1999ns, Litim:1999id, Litim:1999ca}.} would be relevant for a more complete description of a non-abelian plasma from 
kinetic theory. Vlasov equations with spin can be obtained from first principles using Wigner functions and thus it would  be interesting to describe currents with spin. 
See e.g., \cite{Yang:2020hri, Weickgenannt:2020aaf} for a recent application of Wigner functions in the case with collisions. The KMOC formalism has been applied to the spin 
cases in Ref.\cite{Maybee:2019jus}. In principle, the spin case would  require considering classical limits of spinor wavefunctions. Collision functions in kinetic theory
can be obtained from Wigner functions, which can be computed using scattering amplitudes \cite{DeGroot:1980dk}. Perhaps one can study the inclusion of collisions considering classical
limits of scattering amplitudes within  Wigner functions.  

\addsec{Acknowledgements}
We thank Ben Maybee, Donal O'Connell, and Alasdair Ross for many discussions on the KMOC formalism. We thank Ben Maybee and Donal O'Connell for useful comments 
on the manuscript. We thank Fernando T. Brandt for useful correspondence and Luis A. Hernandez for informative discussions. This work was partially supported by the
STFC grant ST/P0000630/1. The author acknowledges financial support from the Open Physics Hub at the Physics and Astronomy Department in Bologna. Our figures were produced with the help of TikZ-Feynman~\cite{Ellis:2016jkw}. Some of the 
calculations in this paper were done with Feyncalc \cite{Mertig:1990an, Shtabovenko:2016sxi, Shtabovenko:2020gxv}.

\appendix

\section{Temperature dependence}
\label{Atemp}
The temperature can be recovered following the procedure in Ref.\cite{Brandt:2015fna}, where the reader can check details. For simplicity, let us consider the 2-point QED example and consider 
Eq.\eqref{forward}. Using our map, we have 
\begin{align}
\Pi^{\mu\nu}(\bp)=\int \frac{\dd^4 k}{(2\pi)^4} \theta(k_0) 2\pi \delta(k^2-m^2) N(k_0) \bar{\mathcal{A}}^{\mu\nu}_2(k, \bp),
\end{align}
which leads to
\begin{align}
\Pi^{\mu\nu}(\bp)= \frac{1}{(2\pi)^3} \int\frac{\dd^3 \mathbf{k} \ N\left(\sqrt{\mathbf{k}^2+m^2}\right)}{2\sqrt{\mathbf{k}^2+m^2}}
\left.\bar{\mathcal{A}}^{\mu\nu}_2(k, \bp)\right|_{k_0=+\sqrt{\mathbf{k}^2+m^2}}.
\end{align}
 Let us define $k_0= |\mathbf{k}| K_0$ with $K_0 >0$ and use spherical coordinates
\begin{align}
\Pi^{\mu\nu}(\bp)=\frac{1}{(2\pi)^3} \int\frac{\dd |\mathbf{k}| \ |\mathbf{k}|^2 \ N\left(
|\mathbf{k}| K_0\right)}{2 |\mathbf{k}| K_0}
\int \dd \Omega
\left.\bar{\mathcal{A}}^{\mu\nu}_2(k, \bp)\right|_{k_0=|\mathbf{k}| K_0}
\end{align}
We can simplify the above expression by introducing the unit vector vector $\hat{\mathbf{K}}=\mathbf{k}/|\mathbf{k}|$ and defining the four vector 
\begin{align}
K^\mu =   (k_0/|\mathbf{k}|, \mathbf{k}/|\mathbf{k}|),
\end{align}
such that $k^\mu= |\mathbf{k}| K^\mu$. Since $\bar{\mathcal{A}}^{\mu\nu}_2(k, \bp)$ is homogeneous function of degree zero in $k$ 
we can then write 
\begin{align}
\Pi^{\mu\nu}(\bp)=\frac{1}{(2\pi)^3} \int_0^\infty\frac{\dd |\mathbf{k}| \ |\mathbf{k}|^2 \ N\left( 
|\mathbf{k}| K_0\right)}{2 |\mathbf{k}| K_0}
\int \dd \Omega
\bar{\mathcal{A}}^{\mu\nu}_2(K, \bp)
\end{align}
Using $K_0= \sqrt{1+ m^2/|k|^2}$ and the change of variables $x=|\mathbf{k}|/T$ leads to 
\begin{align}
\Pi^{\mu\nu}(\bp)= -\frac{1}{(2\pi)^3} 
I(T, m)
\int \dd \Omega\bar{\mathcal{A}}^{\mu\nu}_2(K, \bp),
\end{align} 
where we have included a minus sign to indicate that we are interested in the the classical limit where a fermion is running in the loop.
The temperature dependence is then obtained from 
\begin{align}
I(T, m) \equiv T^2 \int_0^\infty \dd x\frac{x^2}{2 \sqrt{x^2+\frac{m^2}{T^2}} }\ \frac{1}{\exp\left( \sqrt{x^2+\frac{m^2}{T^2}}\right)+1}
\end{align}

\section{3-point QED example}
\label{QED-3pt}
We calculate the 3-point function for QED. The current requires the computation of 12 graphs. We will write the current in terms of the following
bases of tensor structures
\begin{align}
\AC^{\mu_1 \mu_2\mu_3}(k,\bp)=\frac{2 e^3}{(k\cdot \bp_1)^2} \sum\limits_{i=1}^{23} \mathcal{T}_i a_i,
\end{align}
where
\begin{align}
&a_1=\frac{\bar{p}_2{}^2 \bar{p}_1\cdot \bar{p}_3}{\left(k\cdot \bar{p}_2\right){}^2}+\frac{\bar{p}_3{}^2 \bar{p}_1\cdot \bar{p}_2}{\left(k\cdot \bar{p}_3\right){}^2}
-\frac{2 \bar{p}_1\cdot \bar{p}_2 \bar{p}_1\cdot \bar{p}_3}{k\cdot \bar{p}_2 k\cdot \bar{p}_3},\ 
a_2=-\bar{p}_1{}^2,\
a_3=-\frac{\bar{p}_2{}^2 \left(k\cdot \bar{p}_1\right){}^2}{\left(k\cdot \bar{p}_2\right){}^2}, \nonumber\\
&a_4=a_{12}=-k\cdot \bar{p}_1, \
a_5= \frac{\left(k\cdot \bar{p}_1\right){}^2}{k\cdot \bar{p}_2},\ 
a_6=a_7=-\frac{\bar{p}_3{}^2 k\cdot \bar{p}_1}{\left(k\cdot \bar{p}_3\right){}^2}+\frac{\bar{p}_1\cdot \bar{p}_3 k\cdot \bar{p}_1}{k\cdot \bar{p}_2 k\cdot \bar{p}_3}
\nonumber\\
&a_8=\frac{\bar{p}_2{}^2 k\cdot \bar{p}_3}{\left(k\cdot \bar{p}_2\right){}^2}+\frac{\bar{p}_3{}^2}{k\cdot \bar{p}_2}-
\frac{2 \bar{p}_1\cdot \bar{p}_2}{k\cdot \bar{p}_2}, \
a_9= a_{10}=\frac{k\cdot \bar{p}_1}{k\cdot \bar{p}_2}, \
a_ {11}=-\frac{\bar{p}_3{}^2 \left(k\cdot \bar{p}_1\right){}^2}{\left(k\cdot \bar{p}_3\right){}^2},\nonumber\\
&a_ {13}= a_{20}=-\frac{\bar{p}_2{}^2 k\cdot \bar{p}_1}{\left(k\cdot \bar{p}_2\right){}^2}+\frac{\bar{p}_1\cdot \bar{p}_2 k\cdot \bar{p}_1}{k\cdot \bar{p}_2 k\cdot \bar{p}_3},\
a_ {14}= a_{21}=a_{22}= -\frac{\left(k\cdot \bar{p}_1\right){}^2}{k\cdot \bar{p}_2 k\cdot \bar{p}_3}, \
a_ {15}= \frac{k\cdot \bar{p}_1}{k\cdot \bar{p}_2},\nonumber\\
&a_ {16}=\frac{\bar{p}_2{}^2}{k\cdot \bar{p}_3}+\frac{\bar{p}_3{}^2 k\cdot \bar{p}_2}{\left(k\cdot \bar{p}_3\right){}^2}-\frac{2 \bar{p}_1\cdot \bar{p}_3}{k\cdot \bar{p}_3},\
a_ {17}= a_{18}= a_{23}= \frac{k\cdot \bar{p}_1}{k\cdot \bar{p}_3},\
a _{19}=\frac{\left(k\cdot \bar{p}_1\right){}^2}{k\cdot \bar{p}_3}.\nonumber
\end{align}
and
\begin{align}
& \mathcal{T}_1=k^{\mu _1} k^{\mu _2} k^{\mu _3}, \
\mathcal{T}_2= \eta^{\mu_2 \mu_3}  k^{\mu _1},\
\mathcal{T}_3= \eta^{\mu_1 \mu_3 } k^{\mu _2},\
 \mathcal{T}_4=\eta_{\mu_2 \mu_2 } \bar{p}_2{}^{\mu _1}, \
 \mathcal{T}_5=\eta^{\mu_1 \mu_3 } \bar{p}_2{}^{\mu _2}, \nonumber\\
& \mathcal{T}_6=k^{\mu _2} k^{\mu _3} \bar{p}_2{}^{\mu _1},\
 \mathcal{T}_7=k^{\mu _1} k^{\mu _3} \bar{p}_2{}^{\mu _2},\
 \mathcal{T}_8=k^{\mu _1} k^{\mu _2} \bar{p}_2{}^{\mu _3},\
 \mathcal{T}_9=k^{\mu _2} \bar{p}_2{}^{\mu _1} \bar{p}_2{}^{\mu _3},\
 \mathcal{T}_{10}=k^{\mu _1} \bar{p}_2{}^{\mu _2} \bar{p}_2{}^{\mu _3},\nonumber\\
& \mathcal{T}_{11}=\eta^{\mu_1 \mu_2 } k^{\mu _3}, 
 \mathcal{T}_{12}=\eta^{\mu_2 \mu_3 } \bar{p}_3{}^{\mu _1},\
 \mathcal{T}_{13}=k^{\mu _2} k^{\mu _3} \bar{p}_3{}^{\mu _1},\
 \mathcal{T}_{14}=k^{\mu _3} \bar{p}_2{}^{\mu _2} \bar{p}_3{}^{\mu _1},\
 \mathcal{T}_{15}=k^{\mu _2} \bar{p}_2{}^{\mu _3} \bar{p}_3{}^{\mu _1}, \nonumber\\
& \mathcal{T}_{16}=k^{\mu _1} k^{\mu _3} \bar{p}_3{}^{\mu _2},\
 \mathcal{T}_{17}=k^{\mu _3} \bar{p}_2{}^{\mu _1} \bar{p}_3{}^{\mu _2},\
 \mathcal{T}_{18}=k^{\mu _3} \bar{p}_3{}^{\mu _1} \bar{p}_3{}^{\mu _2},\
 \mathcal{T}_{19}=\eta^{\mu_1 \mu_2 } \bar{p}_3{}^{\mu _3},\
 \nonumber\\
&\mathcal{T}_{20}=k^{\mu _1} k^{\mu _2} \bar{p}_3{}^{\mu _3}, \mathcal{T}_{21}=k^{\mu _2} \bar{p}_2{}^{\mu _1} \bar{p}_3{}^{\mu _3},\
 \mathcal{T}_{22}=k^{\mu _1} \bar{p}_2{}^{\mu _2} \bar{p}_3{}^{\mu _3}, \
 \mathcal{T}_{23}=k^{\mu _1} \bar{p}_3{}^{\mu _2} \bar{p}_3{}^{\mu _3}.\nonumber
 \end{align}
We have checked that this result is equivalent to the high temperature limit computed in Ref.\cite{Brandt:2015fna}.The equivalence between these functions and solutions of the Abelian version 
of Eq.\eqref{Vlasov} is discussed Ref.\cite{Brandt:2015fna}.

\bibliographystyle{./JHEP}
 \renewcommand\bibname{References} 
\ifdefined\phantomsection		
  \phantomsection  
\else
\fi
\addcontentsline{toc}{section}{References}


\providecommand{\href}[2]{#2}\begingroup\raggedright\endgroup

\end{document}